\documentclass[aps,prl,twocolumn,showpacs,superscriptaddress,preprintnumbers,floatfix]{revtex4}
\usepackage{amsmath,amsfonts,amssymb,dcolumn,bm}
\usepackage[english]{babel}
\usepackage[dvips]{graphicx}

\begin{document}
\title{Observation of post-soliton expansion following laser propagation\\ through an underdense plasma}

\author{G. Sarri}
\affiliation{School of Mathematics and Physics, Queens University Belfast, BT7 1NN, UK}
\author{D.K. Singh}
\affiliation{GoLP, Instituto de Plasmas e Fus\~{a}o Nuclear - Laborat\'orio Associado, Instituto Superior T\'ecnico, 1049-001 Lisbon, Portugal}
\author{J.R. Davies}
\affiliation{GoLP, Instituto de Plasmas e Fus\~{a}o Nuclear - Laborat\'orio Associado, Instituto Superior T\'ecnico, 1049-001 Lisbon, Portugal}
\author{K.L. Lancaster}
\affiliation{STFC Rutherford Appleton Laboratory, Didcot, OX11 0QX, UK}
\author{E.L. Clark}
\affiliation{Technological Educational Institute of Crete, GR 710 04 Greece}
\author{S. Hassan}
\affiliation{Technological Educational Institute of Crete, GR 710 04 Greece}
\author{J. Jiang}
\affiliation{GoLP, Instituto de Plasmas e Fus\~{a}o Nuclear - Laborat\'orio Associado, Instituto Superior T\'ecnico, 1049-001 Lisbon, Portugal}
\author{N. Kageiwa}
\affiliation{Graduate School of Engineering, Osaka 565-0871, Japan}
\author{N. Lopes}
\affiliation{GoLP, Instituto de Plasmas e Fus\~{a}o Nuclear - Laborat\'orio Associado, Instituto Superior T\'ecnico, 1049-001 Lisbon, Portugal}
\author{A. Rehman}
\affiliation{Blackett Laboratory, Imperial College London, London SW7 2BZ UK}
\author{C. Russo}
\affiliation{GoLP, Instituto de Plasmas e Fus\~{a}o Nuclear - Laborat\'orio Associado, Instituto Superior T\'ecnico, 1049-001 Lisbon, Portugal}
\author{R.H.H. Scott}
\affiliation{STFC Rutherford Appleton Laboratory, Didcot, OX11 0QX, UK}
\affiliation{Blackett Laboratory, Imperial College London, London SW7 2BZ UK}
\author{T. Tanimoto}
\affiliation{Graduate School of Engineering, Osaka 565-0871, Japan}
\author{Z. Najmudin}
\affiliation{Blackett Laboratory, Imperial College London, London SW7 2BZ UK}
\author{K.A. Tanaka}
\affiliation{Graduate School of Engineering, Osaka 565-0871, Japan}
\author{M. Tatarakis}
\affiliation{Technological Educational Institute of Crete, GR 710 04 Greece}
\author{M. Borghesi}
\affiliation{School of Mathematics and Physics, Queens University Belfast, BT7 1NN, UK}
\author{P.A. Norreys}
\affiliation{STFC Rutherford Appleton Laboratory, Didcot, OX11 0QX, UK}
\affiliation{Blackett Laboratory, Imperial College London, London SW7 2BZ UK}

\date{\today}
\begin{abstract}
The expansion of electromagnetic post-solitons emerging from the interaction of a 30 ps, $3\times 10^{18}$ W cm$^{-2}$ laser pulse with an underdense deuterium plasma has been observed up to 100 ps after the pulse propagation, when large numbers of post-solitons were seen to remain in the plasma. The temporal evolution of the post-solitons has been accurately characterized with a high spatial and temporal resolution. The observed expansion is compared to analytical models and three dimensional particle-in-cell results providing indication of the polarisation dependence of the post-soliton dynamics.
\end{abstract}


\maketitle
\begin{figure}[!h]
\begin{center}
\includegraphics[width=8cm,height=4cm]{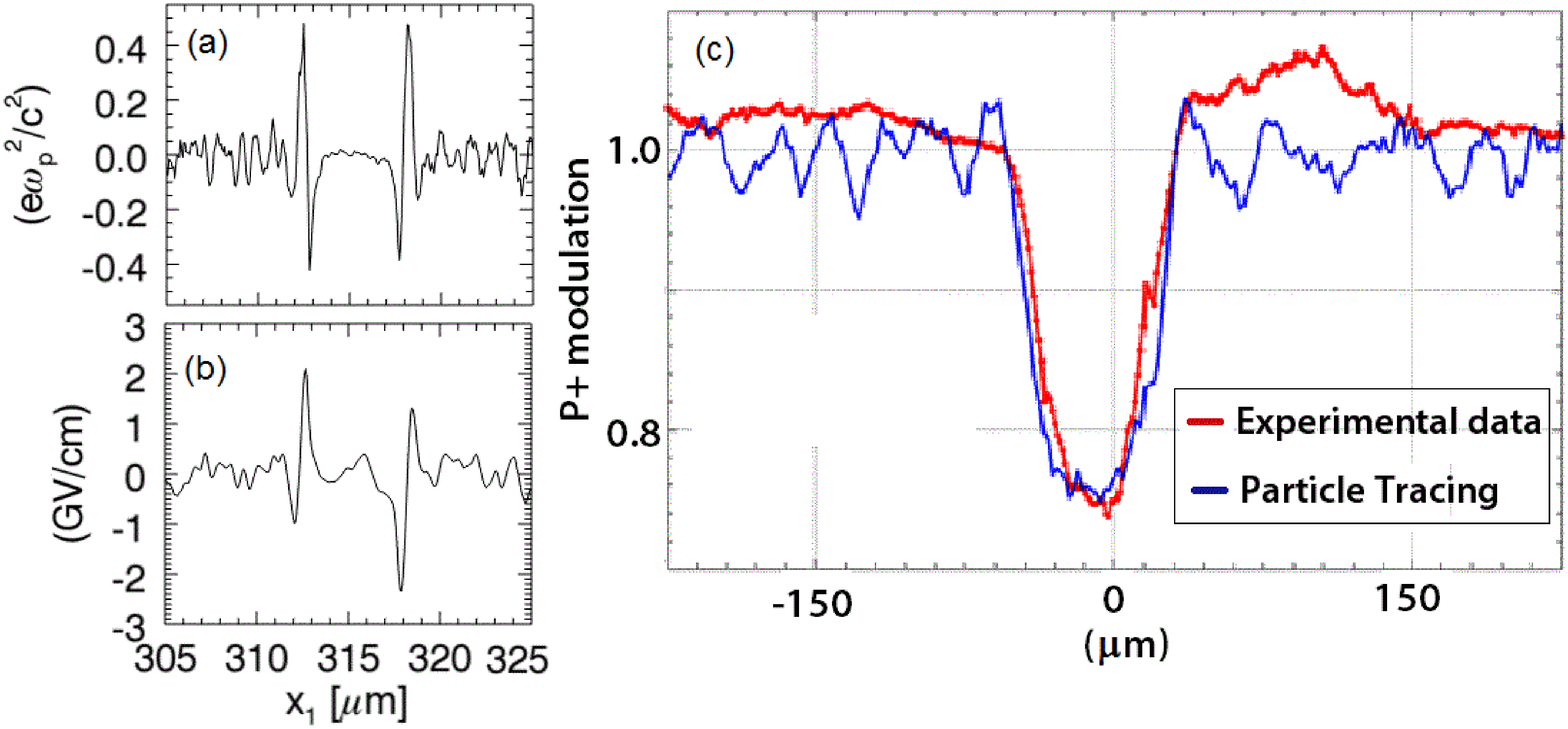}
\caption{\textbf{a)} Cross-cut of the post-soliton charge density from 2D PIC modelling at $3$ ps. \textbf{b)} Corresponding electric field distribution. \textbf{c)} Comparison between the modulation of the proton probing beam, as directly obtained from the RCF (red line) and the modulation simulated via the PT code (blue line).}
\label{particletracer}
\end{center}
\end{figure}

The propagation of laser pulses through underdense plasma is subject to a complex interplay of dispersion effects, due to finite  electron inertia, and nonlinearities, due to the relativistic mass increase of the electrons and particle redistribution by the laser ponderomotive force \cite{Kruer}.
This leads to many well-known non-linear phenomena, including relativistic self-focussing \cite{selffocussing} and soliton formation \cite{Farina, Bulanov, Esirkepov, Lehmann}.\\
\verb|  |During its propagation through an underdense plasma, the laser experiences a significant energy loss. Being this energy loss fully adiabatic, it is mostly translated into a red shift of the laser light \cite{Bulanov2}. In the case of initial plasma densities close to the critical density, this frequency decrease may eventually lead the laser to locally experience an overcritical plasma, thus being trapped in plasma cavities. These cavities, whose radius is of the order of the electron collisionless skin depth ($l_e = c/\omega_{pe}$, where $\omega_{pe}$ is the Langmuir plasma frequency) are usually referred as electromagnetic (e.m.) sub-cycle solitons \cite{Esirkepov}. These structures tend to be accelerated along plasma density gradients \cite{Bulanov,Sentoku} and, therefore, are slowly propagating, if not steady, in a homogeneous plasma. For times longer than the ion plasma period, the Coulomb repulsion of the ions left inside the cavity causes it to radially expand, and the soliton nature is lost: such late-time evolution of a soliton is thus commonly referred as a \emph{post-soliton} \cite{Naumova}.
This mechanism is similar to the Coulomb explosion that occurs inside laser channels following relativistic self-focussing \cite{selffocussing}.\\
\verb|  |Post-soliton expansion has been analytically studied \cite{Naumova,Bulanov3}, using the so-called snowplow model \cite{snowplow} and the isolated spherical resonator model \cite{Landau}, and numerically, using Particle-In-Cell (PIC) codes \cite{Naumova}. Possible experimental indication of the presence of e.m. soliton structures in underdense plasmas has been reported in \cite{Chen} whereas, experimental observation of post-soliton structures was first reported in \cite{Borghesi}, where soliton remnants were observed in the dense region of a plasma resulting from the laser-driven explosion of a thin foil. Due to the nature of the plasma employed, clouds of bubble-like structures were detected thus preventing a precise characterisation of the temporal evolution of the post-solitons.\\
\verb|  |In this Letter, we report the first experimental observation of well isolated post-soliton structures. This allows their evolution to be resolved and followed over a significant temporal window. The experimental results are compared to analytical and three dimensional (3D) PIC code models.\\
\begin{figure}[!h]
\begin{center}
\includegraphics[width=8.5cm,height=3.5cm]{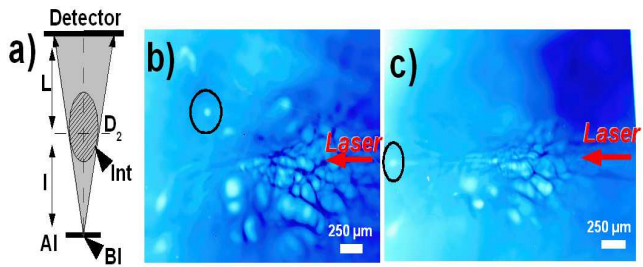}
\caption{\textbf{a)} Top view of the experimental arrangement. \textbf{b), c)}  Radiographs of two different shots: outlined with dark circles are the density bubbles interpreted to be post-solitons.}
\label{setup}
\end{center}
\end{figure}
\verb|  |The experiment was carried out at the Rutherford Appleton Laboratory employing the VULCAN Nd:glass laser operating in the chirped pulse amplification mode \cite{tawest}.
A sketch of the set up is given in Fig.\ \ref{setup}: 200 J of 1 $\mu$m laser light contained in a 30 ps Full Width Half Maximum pulse (``Int" in Fig.\ \ref{setup}) were focussed to a peak intensity of $3\times 10^{18}$ W cm$^{-2}$ at the edge of a supersonic deuterium gas jet with a backing pressure ranging from 1 to 100 bar. This resulted, once fully ionized, in an electron density of $10^{18}$--$10^{20}$ cm$^{-3}$, which is $0.001$--$0.1$ times the non-relativistic critical density $n_c$.
The interaction was diagnosed via the proton radiography technique \cite{Borghesi,Sarri}, which uses, as a particle probe, a laser accelerated proton beam, arising from the interaction of a secondary laser pulse ($\tau\approx 1$ ps, $E\approx 100$ J, $I\geq10^{19}$ W cm$^{-2}$, ``Bl" in Fig.\ \ref{setup}) with a $20\mu$m thick aluminium foil.
The virtual point-like source \cite{Borghesi2} allows imaging of the interaction area with a geometrical magnification $M\approx (l+L)/l\approx6$, where $l\approx6$ mm and $L\approx3$ cm, see Fig.\ \ref{setup}.a.
The probe beam, after having passed through the gas jet, was recorded by a stack of RadioChromic Films (RCF) \cite{Dempsey}.\\
\verb|  |The data set comprised about 30 shots in which both the deuterium density and the probe time were varied. Two typical proton radiographs obtained at a density of $0.1n_c$ are shown in Fig. \ref{setup}.
These images were obtained with protons of energy $\approx4$ MeV, $100$ ps after the beginning of the interaction.
As a rule of thumb, the electric fields are directed from the regions of lighter blue color (reduced proton flux) towards the regions of darker blue color (increased flux).
In both images, a channel created by the laser pulse is visible in the low density region at the edge of the gas jet. In the dense region inside the gas jet a strongly modulated deflection pattern is visible along the laser propagation axis. This scaly region highlights the presence of a cloud of bubbles that appear merged or overlying one another in this 2D projection, possibly surrounding the laser-driven channel. Such a region visually resembles the cloud of solitons that was experimentally and numerically observed in \cite{Borghesi}. Ahead of and around this region, isolated bubble-like structures are visible (black circles in Fig. \ref{setup}), most of them located at the end of laser filaments, as numerically predicted in \cite{Bulanov}. These bubbles are associated with strong probe proton depletion with sharp edges.
We note that these bubble-like structures, which we ascribe to post-solitons, were never observed at electron densities of $0.01n_c$ or less.\\
\verb|  |Considering the isolated bubbles allows following the fundamental properties of the post-solitons temporal evolution.
\begin{figure}[!h]
\begin{center}
\includegraphics[width=7cm,height=7cm]{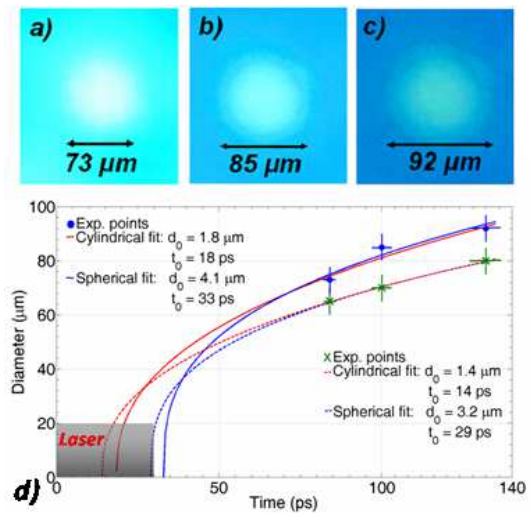}
\caption{Zoom in on the bubble structure outlined in Fig.\ \ref{setup} in different layers of the RCF stack corresponding to 84 ps (\textbf{a)}), 100 ps (\textbf{b)}) and 132 ps (\textbf{c)}). \textbf{d)} Bubble diameter as a function of time and fits using results from the snowplow model for cylindrical and spherical post-solitons.}
\label{bubble_expansion}
\end{center}
\end{figure}
Thanks to the multi-frame capability of proton radiography \cite{Sarri}, it is possible to follow their temporal evolution of these bubbles in the range $80$--$130$ ps after the beginning of the interaction, i.e. $40$--$90$ ps after the laser has left the gas jet.
Both bubbles analyzed were found to be effectively stationary in the laboratory reference frame and to expand preserving a roughly circular shape (Fig.\ \ref{bubble_expansion}).
Bulanov and Pegoraro \cite{Bulanov3} give analytical results for the expansion of 1D planar, 2D cylindrical and 3D spherical post-solitons using the snowplow model.
In 3D the diameter of the sphere is given by $d_0(3^{3/2}t/t_s)^{1/3}$ for $t \gg t_s$, where $t_s$ is given by $\sqrt{2\pi d_0^2n_0m_i/\left\langle E_0^2\right\rangle}$, $d_0$ is the initial diameter, $n_0$ is the initial ion density, $m_i$ is the ion mass and $\left\langle E_0^2\right\rangle$ is the time average of the square of the initial oscillating electric field inside the post-soliton.
We fitted the experimental results with this function taking the initial diameter $d_0$ and the time of creation of the soliton $t_0$ as free parameters and $\left\langle E_0^2\right\rangle^{1/2} = 2\times10^{12}$ V m$^{-1}$, which is roughly the average value of 40\% of the initial laser field. The use of this value for $\left\langle E_0^2\right\rangle$ is justified by recent theoretical results which show that almost 40\% of the initial laser field is trapped within an e.m. soliton \cite{Bulanov}.
Even though this function was able to fit the experimental data (see Fig. \ref{bubble_expansion}), it implied a creation time at the end of, if not after, the laser pulse duration, which is not physically sensible.
We therefore tried the 2D cylindrical result $d_0(5t/t_s)^{2/5}$: this gave a more physically meaningful fit with a creation time close to the peak of the laser pulse in both cases.
The initial diameters from the cylindrical fits are also more reasonable than those from the spherical fits, being $\approx 1$ $\mu$m instead of $\approx 3$ $\mu$m, since we have $c/\omega_{pe} \approx 0.53$ $\mu$m (see Fig. \ref{bubble_expansion} for the fits result). For $d_0\approx 1\mu$m, the time scale of the post-soliton expansion is $t_s\approx 68$fs, justifying the assumption $t-t_0\gg t_s$. Other bubbles in Fig. \ref{setup} and in different shots (not shown) have been found to expand in a similar fashion.\\
\verb|  |In order to understand why a 3D structure follows predictions for 2 rather than 3 dimensions, we carried out a 3D run with the PIC code OSIRIS \cite{Osiris}.
We considered a linearly polarized laser pulse with a wavelength of $1$ $\mu$m, Gaussian spatial and temporal profiles with FWHM of $6$  $\mu$m and $1$ ps, respectively, and a peak intensity of $3 \times 10^{18}$ W cm$^{-2}$ incident on a fully ionized deuterium plasma with a density of $0.1n_c$.
The interaction was followed for $5$ ps.
The simulation box was $350 \times 50 \times 50$ $\mu$m, divided into $2.4 \times 10^8$ cells each having $2$ particles for electrons and $2$ for deuterium ions, the time step was $0.196$ fs.
The pulse duration and time considered are $30$ times less than those in the experiment because of computational limitations.
However, this was long enough for post-soliton formation to occur and to follow their expansion after the passage of the laser pulse.
A number of longer 2D runs, with a larger number of particles per cell and a range of plasma densities, were also carried out for $s$-polarization (laser electric field out of the plane) and $p$-polarization (laser electric field in the plane).
\begin{figure}[!h]
\begin{center}
\includegraphics[width=9cm,height=4.5cm]{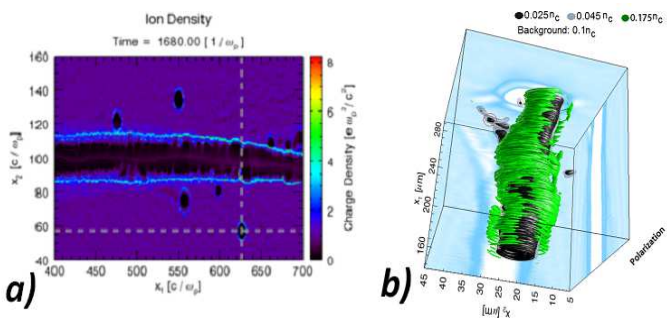}
\caption{\textbf{a)} Ion density at $3$ps from 2D PIC modeling for s-polarized light. The dashed lines highlight the post-soliton from which the electric field distribution in Fig. \ref{particletracer} was taken. \textbf{b)} Iso-surfaces of the ion density at $4.11$ ps from 3D PIC modeling. Sub-channels are formed in the $s$-plane (perpendicular to the laser electric field) following leakage of the laser from the channel formed. Several post-solitons are located in these sub-channels.}
\label{3Dbubbles}
\end{center}
\end{figure}
The ion density in Fig. \ref{3Dbubbles}.b shows prolate spheroid post-solitons, with an aspect ratio of 5:3, lying outside the channel formed by the laser in the plane perpendicular to the laser electric field. In the 2D runs solitons were only ever formed for $s$-polarized light (see, for instance Fig. \ref{3Dbubbles}.a). This dependence of soliton creation upon the laser polarization is in line with the PIC code results reported in \cite{Bulanov}. The solitons were formed as the result of laser leakage due to its breaking-up in filaments that led to the creation of sub-channels departing from the main channel (as visible in both Figs \ref{3Dbubbles}.a and \ref{3Dbubbles}.b). The post-solitons are elongated along the direction of the leaking radiation so that, in principle, they can lead to a circular or elliptical 2D projection in an experimental arrangement as the one shown in Fig. \ref{setup}.a.\\
\verb|  |The non-spherical shape of the post-solitons at these early times gives a first indication as to why the spherical scaling may not apply.
\begin{figure}[!h]
\begin{center}
\includegraphics[width=8cm,height=5.5cm]{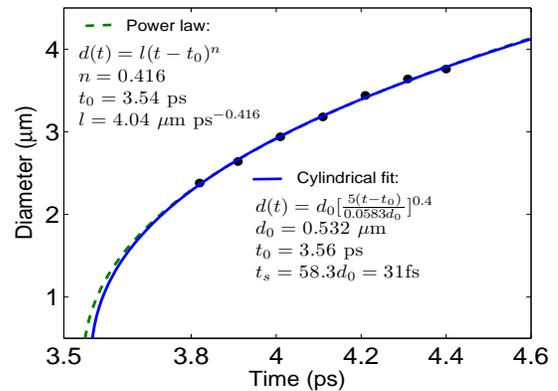}
\caption{Post-soliton diameter X$_2$ extracted from the 3D PIC results along with a power law fit and a fit using the analytical result for a cylindrical post-soliton.}
\label{scaling}
\end{center}
\end{figure}
To determine the scaling of the post-soliton expansion with time, we measured the full width at one quarter of the initial density of the post-soliton circled in Fig.\ \ref{3Dbubbles} along X$_2$ (the smaller diameter) at various times, which is given in Fig.\ \ref{scaling}.
This definition of diameter was chosen because the electric field, which determines the shapes seen by the proton probing, showed a very high noise level (Fig.\ \ref{3Dbubbles}).
However, the precise definition used and the direction in which it is applied did not change the scaling.
We fitted these points with a power law (Fig.\ \ref{scaling}), which gave $t^{0.416}$, clearly in better agreement with the cylindrical scaling of $t^{2/5}$ than the spherical scaling of $t^{1/3}$ as the experimental data suggest.
Following this, we fitted the diameter with the cylindrical scaling of $d_0[5(t-t_0)/t_s]^{2/5}$ taking $t_0$ and $d_0$ as free parameters and $\langle E_0^2 \rangle^{1/2} = 2.4\times10^{12}$ V m$^{-1}$, taken directly from the code results for this post-soliton, which is close to the value assumed in fitting the experimental results.
This gave $t_0=3.5$ ps and $d_0=0.53$ $\mu$m, which gives $t_s = 31$ fs (Fig.\ \ref{scaling}).
Here the assumption of the analytical model that $t-t_0>>t_s$ is again clearly satisfied.
The laser was still within the computational box at $3.5$ ps.
The value of $d_0$ appears somewhat low, since it is less than $2l_e$, but this is due to the way the diameter was determined, which clearly gave a value smaller than the outer edge of the post-soliton, which was difficult to determine unambiguously from the 3D results.\\
\verb|  |The observed polarization dependence of soliton formation suggests why the spherical scaling does not apply. The snowplow model assumes in fact total reflection of the trapped light at the overcritical soliton wall and therefore no plasma heating. This is a good approximation only for $s$-polarized light, since a $p$-polarized wave will be indeed absorbed by the overcritical walls \cite{Wilks}. This might explain why $p$-polarized light is not able to excite stable e.m. solitons \cite{Bulanov}.\\
\verb|  |The electric field structure in the post-soliton was much clearer in the 2D PIC results.
Fig.\ \ref{particletracer} shows a cross-cut through the electric field of the post-soliton outlined by the dashed lines in Fig. \ref{3Dbubbles}.b, and the corresponding charge density.
The electric field shows two sharp, bipolar peaks located at the edge of the post-soliton with maximum amplitude, at 3ps after the beginning of interaction, of $2\times 10^{11}$ V m$^{-1}$; this field would deflect the probe protons in the experiment.
In order to compare this result with the experimental ones, we used a particle tracing (PT) code that traces proton trajectories from a point-like source through a given time-dependent 3D electric field distribution up to the proton detector, giving a 2D proton density map at the detector plane.
The electric field profile used as an input for these simulations was the one produced by the 2D run with a spherically symmetric distribution around its center. The electrostatic field is expected to scale as $E^2 \sim d^{-4}$; this is concluded on the basis that the ratio between the energy and the frequency of the trapped electric field is an adiabatic invariant during the post-soliton expansion \cite{Naumova}. Given a cylindrical scaling for the post-soliton diameter, we expect thus the amplitude of the electric field to decrease in time as $E \sim t^{-4/5}$.
At $100$ ps, the time of the measurements, the maximum amplitude of the electrostatic field should thus be $2\times 10^{11} (100/3)^{-4/5} \approx 10^{10}$ V m$^{-1}$. Such a strong electric field would induce an almost complete depletion of the probing protons in correspondence to the post-soliton structure, as the experimental raw data indicate.
Therefore, the initial conditions for the PT code were a uniform proton beam with an energy of $4$ MeV crossing an electrostatic field with a spatial profile as the one in Fig.\ \ref{particletracer}, with spherical symmetry and a maximum amplitude of $10^{10}$ V m$^{-1}$; the bipolar peaks of the electric field extended for $10\mu$m each, separated by a plateau region 60$\mu$m long.
A cross-cut of the code results and the experimental results is shown in Fig.\ \ref{particletracer}; the two curves show good quantitative agreement, further confirmation of the interpretation of the structures as post-solitons.\\
\verb|  |In summary, we have reported the first experimental measurements of the late time expansion of post-solitons following the propagation of a relativistically intense laser pulse through an underdense plasma. The post-soliton expansion has been temporally resolved with high temporal and spatial resolution. The post-soliton expansion is best described by the analytical prediction for cylindrical, not spherical, post-solitons.
3D PIC code results show the same behavior and, consistently with 2D results, indicate that this is due to polarization effects.\\
\verb|  |The authors would like to warmly thank the staff of the Central Laser Facility for their invaluable help, and the OSIRIS consortium for the use of OSIRIS. This work was carried out under the auspices of the HiPER preparatory project and was supported by the UK Science and Technology Facilities Council.

\end{document}